\documentclass[conference]{IEEEtran}
\IEEEoverridecommandlockouts

\usepackage[utf8]{inputenc}
\usepackage{amsmath}
\usepackage{amssymb}
\usepackage{amsthm}
\usepackage{enumerate}
\usepackage{graphics} 

\usepackage{algorithm}
\usepackage[noend]{algpseudocode}

\usepackage{tikz}
\usetikzlibrary{arrows,shapes,chains,matrix,positioning,scopes,patterns,calc,
decorations.markings,
decorations.pathmorphing,
}

\usepackage{pgfplots}
\pgfplotsset{compat=1.3}
\usepgflibrary{shapes}

\newtheorem{theorem}{Theorem}

\newtheorem{proposition}[theorem]{Proposition}

\newcommand{\cv}{\ensuremath{\mathbf{c}}}

\newcommand{\ev}{\ensuremath{\mathbf{e}}}

\newcommand{\rv}{\ensuremath{\mathbf{r}}}
\newcommand{\sv}{\ensuremath{\mathbf{s}}}

\newcommand{\vv}{\ensuremath{\mathbf{v}}}
\newcommand{\wv}{\ensuremath{\mathbf{w}}}
\newcommand{\xv}{\ensuremath{\mathbf{x}}}
\newcommand{\yv}{\ensuremath{\mathbf{y}}}
\newcommand{\zv}{\ensuremath{\mathbf{z}}}

\newcommand{\zerov}{\ensuremath{\boldsymbol{0}}}

\newcommand{\etav}{\ensuremath{\boldsymbol{\eta}}}

\newcommand{\Am}{\ensuremath{\mathbf{A}}}

\newcommand{\IDm}{\ensuremath{\mathbf{I}}}

\newcommand{\Phim}{\ensuremath{\mathbf{\Phi}}}

\newcommand{\Ncal}{\ensuremath{\mathcal{N}}}
\newcommand{\Ocal}{\ensuremath{\mathcal{O}}}

\title{Multi-User SR-LDPC Codes via Coded Demixing with Applications to Cell-Free Systems}
\author{Jamison R. Ebert, Jean-Francois Chamberland, Krishna R. Narayanan\\
Department of Electrical and Computer Engineering, Texas A\&M University
\thanks{
This material is based upon work supported, in part, by the National Science Foundation (NSF) under Grants CCF-2131106 \& CNS-2148354.}
}

\begin{document}

\maketitle

\begin{abstract}
Novel sparse regression LDPC (SR-LDPC) codes exhibit excellent performance over additive white Gaussian noise (AWGN) channels in part due to their natural provision of shaping gains. 
Though SR-LDPC-like codes have been considered within the context of single-user error correction and massive random access, they are yet to be examined as candidates for coordinated multi-user communication scenarios. 
This article explores this gap in the literature and demonstrates that SR-LDPC codes, when combined with coded demixing techniques, offer a new framework for efficient non-orthogonal multiple access (NOMA) in the context of coordinated multi-user communication channels. 
The ensuing communication scheme is referred to as MU-SR-LDPC coding.
Empirical evidence suggests that, for a fixed SNR, MU-SR-LDPC coding can achieve a target bit error rate (BER) at a higher sum rate than orthogonal multiple access (OMA) techniques such as time division multiple access (TDMA) and frequency division multiple access (FDMA).
Importantly, MU-SR-LDPC codes enable a pragmatic solution path for user-centric cell-free communication systems with (local) joint decoding.
Results are supported by numerical simulations.
\end{abstract}


\section{Introduction}
\label{section:introduction}

Existing wireless systems are predominantly tailored to human-centric data consumption for applications such as Internet browsing, navigation, and media streaming.
Indeed, many contemporary wireless systems offer efficient and high-throughput downlink capacity through the \emph{acquisition-estimation-scheduling} paradigm.
Two notable shifts in wireless network usage emphasize the necessity for a more robust and resilient uplink.
The first trend is exemplified by the wide adoption of mobile video conferencing and the emergence of content creators live-streaming from wireless devices, which both require substantial uplink throughput.
The second transformation involves machine-type communication, which encompasses applications like federated learning, data acquisition, and remote sensing.
Evolving data trends pose a heightened demand on the uplink, challenging current wireless systems.
Thus, there is a growing research emphasis on efficient and resilient uplink connectivity to prepare infrastructures for future traffic~\cite{ding2017survey,ngo2024ultra}.
The cell-free architecture with distributed Multiple Input Multiple Output (MIMO) signal processing has been proposed as a pragmatic approach to building resilient and scalable infrastructures~\cite{bjornson2020scalable,demir2021foundations}.
Along these lines, this article introduces a novel coding scheme for coordinated multi-user communications that offers excellent performance, enables efficient joint decoding, and is seamlessly applicable to the cell-free setting. 
The envisioned paradigm builds on sparse regression LDPC (SR-LDPC) codes~\cite{ebert2023srldpc}, a recently proposed single-user scheme, and draws inspiration from new results on approximate message passing (AMP) applied to cell-free settings~\cite{ccakmak2023inference}.

SR-LDPC codes are a novel class of concatenated codes consisting of an inner sparse regression code (SPARC)~\cite{joseph2012least,barbier2014amp,venkataramanan2019sparse} and an outer non-binary LDPC code~\cite{bennatan2006design}, where the section size of the SPARC and the field size of the LDPC code are equal. 
SR-LDPC codes may be efficiently decoded via an AMP-based algorithm that runs belief propagation (BP) on the factor graph of the outer LDPC code within each denoising step.
This decoding scheme is termed the AMP-BP algorithm. 
SR-LDPC codes have been shown to match or exceed the performance of highly optimized 5G LDPC codes with bit interleaved coded modulation (BICM) over AWGN channels for parameters of practical interest. 
Moreover, as each SR-LDPC coded symbol is an i.i.d.\ Gaussian random variable, such codes intrinsically inherit a \emph{shaping gain} which enhances performance over AWGN channels and beyond.

SR-LDPC codes are closely related to a specific coded compressed sensing (CCS) scheme tailored to unsourced random access (URA)~\cite{amalladinne2022ccs}.
In URA, the objective is to efficiently support a massive number of uncoordinated users with short packet lengths and sporadic activity patterns~\cite{polyanskiy2017perspective}.
A key aspect of the URA strategy is that all users employ identical codebooks for message transmission.
The aforementioned CCS algorithm proceeds by having each active user generate its own SR-LDPC-like codeword using a common sensing matrix and then by having all active users simultaneously transmit their codewords over the Gaussian multiple access channel (GMAC). 
The receiver then recovers an unordered list of transmitted messages by performing noisy $K$-sparse recovery via a modified AMP-BP algorithm, where $K$ denotes the number of active users. 
Thus, the strategy of concatenating an inner sparse regression code with an outer non-binary LDPC code and decoding using the AMP-BP algorithm has proven effective in both the single-user and massive random access scenarios.
However, this approach has not been investigated in the coordinated multi-user communication scenario, which conceptually lies between these two extremes. 

A fundamental building block that facilitates the development of MU-SR-LDPC codes is coded demixing~\cite{ebert2022codeddemixing}.
While studying CCS for URA, we discovered that performance can be improved by (randomly) partitioning active users into different sparse domains in a manner akin to demixing~\cite{mccoy2014sharp,zhou2017demixing}.
Specifically, the superimposed signals to be recovered should be block sparse with respect to separate domains exhibiting low cross-coherence, and the sections of each signal must be coupled via an outer non-binary LDPC code. 
Under such circumstances, the joint recovery of multiple sparse vectors through their superimposed measurements is conducive to the application of AMP-BP decoding with minimal modifications or added complexity.
A key observation of this paper is that this latter approach is readily applicable to a set of SR-LDPC codes because the ensuing scheme naturally satisfies the requirements of coded demixing.
This idea forms the basis of the proposed MU-SR-LDPC coding scheme.

Interestingly, MU-SR-LDPC codes are suitable for user-centric cell-free communication systems~\cite{bjornson2020scalable,demir2021foundations}.
In traditional cellular systems, a geographic service area is partitioned into cells, where each cell is served by a single base station. 
Due to the physics of electromagnetic propagation and the realities of inter-cell interference, users who are physically located close to the base station enjoy significantly stronger channels than users who are consigned to cell edges.
To promote a more uniform quality of service throughout the coverage area, the cell-free paradigm suggests removing cell boundaries and distributing many access points (APs) throughout the geographic region. 
Therein, active users each attach to the subset of APs for which they have the strongest channels. 
These APs are often connected together through common CPUs and may thus collaborate on uplink and downlink signal processing tasks.
Interestingly, MU-SR-LDPC codes seem naturally applicable to both cellular and cell-free scenarios with minimal algorithmic modifications.
This aligns with the findings recently reported by \c{C}akmak, et al.~\cite{ccakmak2023inference}, albeit with a different coded structure.

\subsubsection{Organization and Main Contributions}
In Section~\ref{section:multi_user_single_cell_model}, we introduce MU-SR-LDPC codes within the context of a single-cell network. 
We then highlight how MU-SR-LDPC coding enables an increased sum-rate at a fixed $E_b/N_0$ value when compared to orthogonal multiple access techniques such as TDMA or FDMA. 
Then, in Section~\ref{section:cell_free_system_model}, we consider the cell-free system model and highlight how MU-SR-LDPC codes may, with minimal modification, be applied to such scenarios. 
In both sections, results are supported by numerical simulations.

\subsubsection{Notation}
Matrices are denoted by bold capital letters such as $\Am$.
Vectors are represented by bold lower case letters such as $\xv$, and scalar entries have the form $x_i$.
Furthermore, vector $\ev_{i}$ is a standard basis vector composed of all zeros, except for a one at location $i$. 
The $\ell_p$-norm of $\xv$ is $\| \xv \|_{p}$.
The short form $[N]$ denotes the set of integers $\{ 0, 1, \ldots, N-1 \}$.

\section{Multi-User Single-Cell Model}
\label{section:multi_user_single_cell_model}

The first setting we consider is the coordinated single-antenna Gaussian multiple access channel (GMAC) where the received signal $\yv \in \mathbb{R}^{n}$ is given by
\begin{equation} \label{eq:mu_system_model}
\textstyle  \yv = \sum_{k \in [K]} \xv_k + \zv .
\end{equation}
Parameter $K$ represents the total number of users, $\xv_k$ denotes the signal transmitted by user~$k$, and $\zv \sim \Ncal\left(\zerov, \sigma^2\IDm\right)$ is additive noise.
The base station is tasked with recovering messages $\{\xv_k : k \in [K]\}$ from the observed signal $\yv$.
Every sent signal is subject to an average power constraint $\mathbb{E}\left[\|\xv\|_2^2\right] \leq C$, where $n$ is the number of channel uses (real degrees of freedom).
As mentioned above, we seek to evaluate the performance of a multi-user access scheme where every user employs an SR-LDPC concatenated code.

LDPC codes~\cite{gallager1962low,richardson2008modern} and sparse regression codes~\cite{joseph2012least,venkataramanan2019sparse} are well established and documented.
Code concatenation~\cite{forney1965concatenated} and, more specifically, the concatenation of sparse regression and LDPC codes~\cite{liang2020compressed,amalladinne2022ccs,ebert2023srldpc} are also well understood.
Thus, we only briefly discuss the individual components and the structure of the proposed coding scheme, focusing instead on the specifics of our problem and empirical performance.
We refer the reader to~\cite{ebert2023srldpc,ebert2022codeddemixing} for thorough treatments of SR-LDPC codes and coded demixing, respectively.

\subsection{MU-SR-LDPC Encoding}

In an MU-SR-LDPC system, the encoding operation is analogous for all active users.
We can therefore focus on the encoding process for a generic user and, for ease of presentation, momentarily neglect the subscript~$k$.
The encoding process begins by taking binary sequence $\wv$ and encoding it into a non-binary LDPC codeword
\begin{equation}
\vv = \left( v_0, \ldots, v_{L-1} \right) \in \mathbb{F}_{q}^{L} ,
\end{equation}
where $L$ is the code length, or the number of $\mathbb{F}_{q}$ symbols.
We use a field of cardinality $q = 2^p$, $p \in \mathbb{N}$, which is especially convenient for encoding binary sequences.
Every symbol $v_{\ell}$ in $\vv$ is mapped to a standard basis vector $\ev_{\phi(v_{\ell})}$ through \emph{indexing}, a.k.a.\ {one-hot encoding}.
Here, $\phi(v): \mathbb{F}_{q} \mapsto \{ 0, \ldots q-1 \}$ is a bijection that maps $0 \in \mathbb{F}_{q}$ to zero and $1 \in \mathbb{F}_{q}$ to one.
The ensuing basis vectors are then stacked together in the following manner:
\begin{equation}
    \label{eq:indexed_fq_codeword}
    \sv = \begin{bmatrix}
        \ev_{\phi(v_0)} \\
        % \ev_{\phi(v_{\ell})} \\
        \vdots \\
        \ev_{\phi(v_{L-1})}
    \end{bmatrix}.
\end{equation}
Accordingly, $\sv \in \mathbb{R}^{qL}$ is a block sparse vector consisting of $L$ one-sparse sections, each of length $q$.
An SR-LDPC codeword $\xv$ is then obtained by pre-multiplying vector $\sv$ by sensing matrix $\Am \in \mathbb{R}^{n \times qL}$, with
\begin{equation}
    \label{eq:srldpc_encoding}
    \xv = \Am\sv .
\end{equation}
Matrix $\Am$ is the outcome of a random experiment where each entry is generated i.i.d. according to $a_{i, j} \sim \Ncal\left( 0, 1/{n} \right)$.
Importantly, the sensing matrices are drawn independently across users, and they remain fixed throughout during transmission.

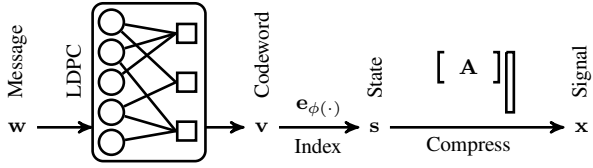
\begin{figure}[t]
\centering
\begin{tikzpicture}
  [
  font=\footnotesize, >=stealth', line width=1pt,
  check/.style={rectangle, minimum height=2.5mm, minimum width=2.5mm, draw=black},
  varnode/.style={circle, minimum size=2mm, draw=black},
  mmse/.style={rectangle, minimum height=7.5mm, minimum width=25mm, rounded corners, draw=black},
  quantity/.style={rectangle, minimum height=8mm, minimum width=8mm, rounded corners, draw=black},
  multiply/.style={trapezium, trapezium angle=75, draw=black, minimum width=10mm, minimum height=8mm, rounded corners}
  ]

\node[rotate=90,anchor=west] (message) at (-0.25,0.25) {Message};
\node (bits) at (-0.25,0) {$\wv$};

\node[rotate=90,anchor=west] (ldpc) at (0.5,0.25) {LDPC};
\draw[draw=black, rounded corners] (0.75, -0.45) rectangle (2.25,1.65);
\foreach \s in {1,2,3,4,5} {
  \node[varnode] (var-\s) at (1,0.4*\s-0.6) {};
}
\foreach \c in {1,2,3} {
  \node[check] (check-\c) at (2,0.65*\c-0.7) {};
}
\draw (var-5) -- (check-3.west);
\draw (var-4) -- (check-3.west);
\draw (var-3) -- (check-3.west);
\draw (var-2) -- (check-2.west);
\draw (var-5) -- (check-2.west);
\draw (var-4) -- (check-1.west);
\draw (var-2) -- (check-1.west);
\draw (var-1) -- (check-1.west);

\foreach \v in {5.75} {
  \draw[line width=1pt] (\v-0.325,1) -- (\v-0.4,1) -- (\v-0.4,0.6) -- (\v-0.325,0.6);
  \draw[line width=1pt] (\v+0.325,1) -- (\v+0.4,1) -- (\v+0.4,0.6) -- (\v+0.325,0.6);
  \draw[line width=1pt] (\v+0.5,1) -- (\v+0.5,0.2) -- (\v+0.6,0.2) -- (\v+0.6,1) -- (\v+0.5,1);
  \node at (\v,0.8) {$\Am$};
}

\node[rotate=90,anchor=west] (codeword) at (3,0.25) {Codeword};
\node (symbols) at (3,0) {$\vv$};

\node (indexing) at (3.75,-0.25) {Index};

\node[rotate=90,anchor=west] (state) at (4.5,0.25) {State};
\node (index) at (4.5,0) {$\sv$};

\node (compression) at (5.75,-0.25) {Compress};

\node[rotate=90,anchor=west] (signal) at (7.25,0.25) {Signal};
\node (x) at (7.25,0) {$\xv$};

\draw[->] (bits) to (0.75,0);
\draw[->] (2.25,0) to (symbols);
\draw[->] (symbols) to node[above]{$\ev_{\phi(\cdot)}$} (index);
\draw[->] (index) to (x);

\end{tikzpicture}
\caption{This depicts the steps of the SR-LDPC encoding process.
Information bits are encoded, the symbols of the LDPC codeword are indexed and stacked, and the resulting vector is compressed through matrix multiplication.
The encoding is analogous for every user, each with their own sensing matrix.}
\label{fig:Encoding}
\end{figure}

We emphasize that, in MU-SR-LDPC systems, each user's unique sensing matrix must be known to both the user and the base station.
This requires some level of coordination or shared randomness.
This assumption is a strong distinction with SR-LDPC codes used in URA settings~\cite{amalladinne2022ccs,ebert2022codeddemixing}, where users are uncoordinated.
However, it is aligned with traditonal MAC settings where every user is assigned its own codebook, spreading sequence, time slot, or frequency tone.
Also, we note that $\xv_i \sim \Ncal\left(0, L/{n}\right)$, therefore channel inputs are i.i.d.\ Gaussian random variables. 
This input distribution is suitable for AWGN channels, and this property can be loosely interpreted as the code benefiting from an inherent \emph{shaping gain} compared to traditional modulation.
A notional diagram for the encoding schemes appears in Fig.~\ref{fig:Encoding}.

\subsection{MU-SR-LDPC Decoding via Coded Demixing}

Having established the encoding process for every active user, we turn to decoding at the access point.
To differentiate users, we must revert back to subscript notation wherever appropriate.
Recall that the structure of the observation signal $\yv$ is additive and, owing to the encoding process, it can be rewritten as
\begin{equation} \label{eq:extended_rx_signal}
\textstyle \yv = \sum_{k \in [K]} \Am_k \sv_k + \zv.
\end{equation}
Conceptually, one may reformulate the expression above by horizontally stacking the $\{\Am_k\}$ matrices to obtain 
\begin{equation}
\Phim = \begin{bmatrix} \Am_0 & \cdots & \Am_{K-1} \end{bmatrix},
\end{equation}
where $\Phim \in \mathbb{R}^{n \times qLK}$.
Likewise, by vertically stacking the $\{\sv_k\}$ vectors, it follows that 
\begin{equation}\label{eq:revised_rx_signal}
\yv = \Phim \begin{bmatrix} \sv_0 \\ \vdots \\ \sv_{K-1} \end{bmatrix} + \zv
= \Phim \mathfrak{s} + \zv .
\end{equation}
We emphasize that \eqref{eq:revised_rx_signal} assumes the canonical form of a compressed sensing problem, with a (very) wide sensing matrix $\Phim$ multiplying stacked vector $\mathfrak{s}$.
This problem can be interpreted as an inverse problem where the base station is tasked with recovering a (structured) sparse vector $\mathfrak{s}$ from noisy observation $\yv$.
It is then intuitive to try to solve this compressed sensing problem through the AMP-BP algorithm, if perhaps at an unprecedented scale.
The standard AMP composite algorithm applied to our stacked problem is:
\begin{align}
\zv^{(t)} &= \yv - \Phim \mathfrak{s}^{(t)} + \frac{\zv^{(t-1)}}{n} \operatorname{div} \mathfrak{H}^{(t-1)} \left( \mathfrak{r}^{(t-1)} \right) \label{equation:AMP-Residual} \\
\mathfrak{r}^{(t)} &= \Phim^{\mathrm{T}} \zv^{(t)} + \mathfrak{s}^{(t)} \label{equation:EffectiveObservation} \\
\mathfrak{s}^{(t+1)} &= \mathfrak{H}^{(t)} \left( \mathfrak{r}^{(t)} \right) . \label{equation:AMP-Denoising}
\end{align}
Superscript $t$ denotes the iteration count and the algorithm is initialized with $\mathfrak{r}^{(0)} = \mathfrak{s}^{(0)} = \zerov$ and $\zv^{(0)} = \yv$.
Furthermore, every quantity with a negative iteration count is equal to the zero vector.

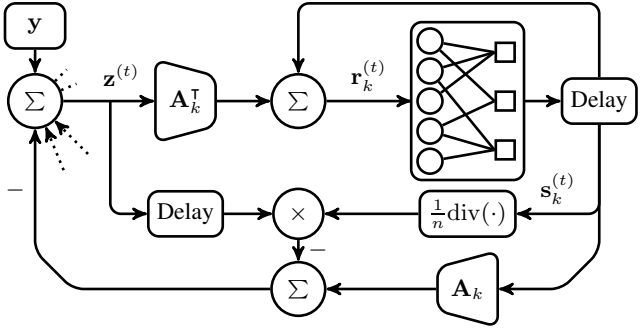
\begin{figure}[t]
\centering
\begin{tikzpicture}
  [
  font=\small, >=stealth', line width=1pt,
  check/.style={rectangle, minimum height=2.5mm, minimum width=2.5mm, draw=black},
  varnode/.style={circle, minimum size=1.5mm, draw=black},
  quantity/.style={rectangle, minimum height=6mm, minimum width=8mm, rounded corners, draw=black},
  multiply/.style={trapezium, trapezium angle=75, draw=black, minimum width=10mm, minimum height=8mm, rounded corners}
  ]

% Received Signal
%
\node[quantity] (signal) at (-4.5,3) {$\mathbf{y}$};

% Residual
%
% \node[quantity] (residual) at (-5.5,2) {};
\node[circle, minimum width=6mm, draw=black] (sum) at (-4.5,2) {\scriptsize $\sum$}
  edge[<-] (signal)
  edge[<-,dotted] (-4.5+0.71,2-0.71)
  edge[<-,dotted] (-4.5+0.38,2-0.92)
  edge[dotted] (-4.5+0.42,2+0.42)
  edge[dotted] (-4.5+0.55,2+0.23);
\node at (-3.35,2.35) {$\mathbf{z}^{(t)}$};

% % % % % % % % % %

\foreach \g/\xg/\yg in {1/1/0} {  % 3/3.5/2,2/2.25/1,
    % Effective Observation
    %
    \node[multiply, shape border rotate=270] (dual\g) at (-3.5+\xg,2+\yg) {$\Am_k^\intercal$};
    \node[multiply, shape border rotate=90] (primal\g) at (0.25+\xg,-0.5+\yg) {$\Am_k$};
    \draw[->, rounded corners] (sum.east) -- (-4.5+\xg,2+\yg) -- (dual\g);
    \node[circle, minimum width=6mm, draw=black] (sum\g) at (-2+\xg,2+\yg) {\scriptsize $\sum$}
      edge[<-] (dual\g);
    \draw[->] (sum\g) -- node[above]{$\rv_k^{(t)}$} (-0.5+\xg,2+\yg);
    \node[quantity] (z\g-delay) at (-3.5+\xg,0.5+\yg) {Delay};
    \node[circle, minimum width=6mm, draw=black] (times\g) at (-2+\xg,0.5+\yg) {$\times$};
    \node[quantity] (div\g) at (0.25+\xg,0.5+\yg) {$\frac{1}{n} \mathrm{div} (\cdot)$}
        edge[->] (times\g);
    \draw[->,rounded corners] (-4.5+\xg,2+\yg) -- (-4.5+\xg,0.5+\yg) -- (z\g-delay);
    \draw[->,rounded corners] (z\g-delay) -- (times\g);
    \node[circle, minimum width=6mm, draw=black] (finalsum\g) at (-2+\xg,-0.5+\yg) {\scriptsize $\sum$}
        edge[<-] (primal\g)
        edge[<-] node[right] {$-$} (times\g);
    
    % Denoiser
    %
    % \draw[draw=black, densely dotted, rounded corners, fill=white] (1+\xg,0.75+\yg) rectangle (4+\xg,2.75+\yg);
    % \node[draw=none] (denoiser\g) at (2.5+\xg,3+\yg) {Denoiser};
    % \node[mmse] (mmse\g) at (2.5+\xg,-0.125+\yg){BP Denoising};
    \draw[draw=black, rounded corners] (-0.5+\xg,0.95+\yg) rectangle (1+\xg,3.05+\yg);
    \foreach \s in {1,2,3,4,5} {
      \node[varnode] (var\g-\s) at (-0.25+\xg,0.8+0.4*\s+\yg) {};
    }
    \foreach \c in {1,2,3} {
      \node[check] (check\g-\c) at (0.75+\xg,0.7+0.65*\c+\yg) {};
    }
    \draw (var\g-5) -- (check\g-3.west);
    \draw (var\g-4) -- (check\g-3.west);
    \draw (var\g-3) -- (check\g-3.west);
    \draw (var\g-2) -- (check\g-2.west);
    \draw (var\g-5) -- (check\g-2.west);
    \draw (var\g-4) -- (check\g-1.west);
    \draw (var\g-2) -- (check\g-1.west);
    \draw (var\g-1) -- (check\g-1.west);
    
    % State Update
    %
    \node[quantity] (state\g-delay) at (2+\xg,2+\yg) {Delay}
        edge[<-] (1+\xg,2+\yg);
    \draw[->,rounded corners] (state\g-delay) -- (2+\xg,3.3+\yg) -- (-2+\xg,3.3+\yg) -- (sum\g);
    \draw[->,rounded corners] (state\g-delay) -- (2+\xg,0.5+\yg) -- node[above] {$\sv_k^{(t)}$} (div\g);
    \draw[->,rounded corners] (state\g-delay) -- (2+\xg,0+\yg) -- (1.5+\xg,-0.5+\yg) -- (primal\g);
    \draw[->,rounded corners] (finalsum\g) -- (-4,-0.5+\yg) -- (-4.5,0+\yg) -- node[left] {$-$} (sum);
}

\end{tikzpicture}
\caption{
    This notional diagram depicts the operation of the MU-SR-LDPC decoder.
    The contribution of the estimated state vector enhanced by the Onsager is computed separately for each user (only one block shown).
    The residual is computed based on the contributions associated with all the users.
    }
\label{fig:Decoding}
\end{figure}

The denoising function $\mathfrak{H}^{(t)} (\cdot)$ we wish to utilize is a variation of the BP denoiser introduced in~\cite{ebert2023srldpc}.
More specifically, the denoiser is separable across users and it operates on individual codeword sections,
\begin{equation*}
\mathfrak{H}^{(t)} \left( \mathfrak{r}^{(t)} \right)
= \begin{bmatrix} \etav^{(t)}_0 \left( \rv_0^{(t)} \right) \\ \vdots \\ \etav^{(t)}_{K-1} \left( \rv_{K-1}^{(t)} \right) \end{bmatrix},
\end{equation*}
where we have implicitly use the correspondences
\begin{xalignat*}{2}
\mathfrak{r}^{(t)}
&= \begin{bmatrix} \rv_0^{(t)} \\ \vdots \\ \rv_{K-1}^{(t)} \end{bmatrix}
= \begin{bmatrix} \Am_0^{\mathrm{T}} \zv^{(t)} + \sv_0^{(t)} \\ \vdots \\ \Am_{K-1}^{\mathrm{T}} \zv^{(t)} + \sv_{K-1}^{(t)} \end{bmatrix}; &
\mathfrak{s}^{(t)} = \begin{bmatrix} \sv_0^{(t)} \\ \vdots \\ \sv_{K-1}^{(t)} \end{bmatrix} .
\end{xalignat*}
Due to space limitations, we cannot explain all the details of the BP denoiser $\etav^{(t)}_k \big( \rv_k^{(t)} \big)$, but the algorithm and its intricacies can be found in~\cite{ebert2023srldpc}.
Still, we provide a conceptual explanation for the operation of $\etav^{(t)}_k \big( \rv_k^{(t)} \big)$.
The user-specific effective observation $\rv_k^{(t)}$ and the user-specific state estimate $\sv_k^{(t)}$ are partitioned into their respective sectional representations,
\begin{xalignat*}{2}
\rv_k^{(t)} &= \begin{bmatrix} \rv_{k,0}^{(t)} \\ \vdots \\ \rv_{k,L-1}^{(t)} \end{bmatrix} ; &
\sv_k^{(t)} &= \begin{bmatrix} \sv_{k,0}^{(t)} \\ \vdots \\ \sv_{k,L-1}^{(t)} \end{bmatrix} .
\end{xalignat*}
Each section $\rv_{k,\ell}^{(t)}$ in $\rv_k^{(t)}$ acts as a vector observation about the true value of $\sv_{k, \ell} \in \left\{ \ev_g : g \in \phi(\mathbb{F}_q) \right\}$.
An interesting property of AMP is that, under certain technical conditions, the effective observation $\rv_{k,\ell}^{(t)}$ is asymptotically distributed as the true basis vector $\sv_{k, \ell}$ plus i.i.d.\ Gaussian noise of known variance.
This property hinges on the presence of the Onsager term in \eqref{equation:AMP-Residual} and some smoothness conditions on the denoiser.
Sectional effective observation $\rv_{k,\ell}^{(t)}$ then induces a probability distribution on the symbols of $\mathbb{F}_q$.
Denoting the variance of the effective noise by $\tau$, we can write
\begin{equation*}
\begin{split}
\boldsymbol{\alpha}_{k, \ell} (g)
&= \Pr \left( \vv_{k, \ell} = g \Big| \rv_{k,\ell}^{(t)} \right)
= \frac{e^{- \frac{\left\| \rv_{k, \ell} - \ev_{\phi(g)} \right\|^2}{2 \tau^2}}}{\sum_{h \in \mathbb{F}_q} e^{- \frac{\left\| \rv_{k, \ell} - \ev_{\phi(h)} \right\|^2}{2 \tau^2}}} .
% = \frac{e^{\frac{\rv_{k,\ell}^{(t)} \left( \phi(g) \right)}{\tau^2}}}
% {\sum_{h \in \mathbb{F}_q} e^{\frac{\rv_{k,\ell}^{(t)} \left( \phi(h) \right)}{\tau^2}}} .
\end{split}
\end{equation*}
Local estimates on $\vv_{k, \ell}$ can then be refined efficiently, leveraging the properties of the outer code by applying standard belief propagation (BP) for non-binary LDPC codes.
The output of the denoiser is taken to be, section-wise, the estimate afforded by BP after a prescribed number of message passing iterations,
\begin{equation*}
\left[ \sv_{k,\ell}^{(t+1)} \right]_i = \left[ \etav^{(t)}_{k} \left( \rv_{k}^{(t)} \right) \right]_{\ell,i}
= \operatorname{BP} \left( \vv_{k, \ell} = \phi^{-1}(i) \Big | \rv_{k}^{(t)} \right) .
\end{equation*}
Admittedly, this description may feel terse; details are omitted due to space.
The denoising function is user separable, but it is not section separable; in other words, $\etav^{(t)}_{k}(\cdot)$ is a non-separable function.
This is to be expected because performing BP iteration on the factor graph of the LDPC code requires estimates of all the sections involved.
Still, it is shown in~\cite{ebert2023srldpc} that BP denoising is Lipschitz continuous under reasonable conditions.
Furthermore, under mild conditions, the divergence operator of the per-user denoising function admits an elegant closed form expression.
Since, the denoiser is user separable, we can write
\begin{equation*}
\begin{split}
\operatorname{div} & \mathfrak{H}^{(t)} \left( \mathfrak{r}^{(t)} \right)
= \sum_{k \in [K]} \operatorname{div} \etav^{(t)}_{k} \left( \rv_k^{(t)} \right) \\
&= \sum_{k \in [K]} \frac{1}{\tau^2} \left( \left\| \etav_k^{(t)} \left( \rv^{(t)}_k \right) \right\|_1 - \left\| \etav_k^{(t)} \left( \rv_k^{(t)} \right) \right\|_2^2 \right) .
\end{split}
\end{equation*}
The exact form of the Onsager term offers limited insight.
However, its additive structure and the fact that the denoiser is separable across users are crucial to our analysis.
Combining these two attributes, we can rewrite the AMP equations as
\begin{align}
\hat{\yv}_k^{(t)} = \Am_k \sv_k^{(t)} &- \frac{1}{n} \zv^{(t-1)} \operatorname{div} \etav_{k}^{(t-1)} \left( \rv_k^{(t-1)} \right) \label{eq:EstimatedContributions} \\
\zv^{(t)} &= \textstyle \yv - \sum_{k \in [K]} \hat{\yv}_k^{(t)} \label{eq:AMP-Residual-Plus} \\
% \zv^{(t)} = \yv - \sum_{k \in [K]} \left( \Am_k \sv_k^{(t)} - \frac{\zv^{(t-1)}}{n} \operatorname{div} \etav_{k}^{(t-1)} \left( \rv_k^{(t-1)} \right) \right) \\
\rv_{k}^{(t)} &= \Am_k^{\mathrm{T}} \zv^{(t)} + \sv_k^{(t)} \label{eq:EffectiveObservation-Plus} \\
\sv_{k}^{(t+1)} &= \etav_{k}^{(t)} \left( \rv_{k}^{(t)} \right) \label{eq:AMP-Denoising-Plus} .
\end{align}
Equation~\eqref{eq:EstimatedContributions} introduces the estimated contribution from user~$k$.
A possible interpretation of the algorithm is that users are jointly trying to best explain their own contributions to observation $\yv$.
We emphasize that the algorithm is highly parallelizable.
To update the state estimate and the Onsager term for user~$k$, all that is needed is $\zv^{(t)}$ and the previous version of the estimate $\sv_k^{(t)}$.
The per user state update balances the current state estimate, the unexplained portion of the observation, and the graphical constraints associated with the outer LDPC code.
An update of the residual only requires the computation of one vector $\hat{\yv}_k^{(t)}$ per user.
We gather that the computational complexity per AMP iterations is linear in the number of users within the cell, or $\Ocal\left(K\right)$.
Fig.~\ref{fig:Decoding} offers a conceptual blueprint for the decoding process.

\subsection{Empirical Results for Single-Cell Setting}

In~\cite{ebert2023srldpc}, we investigate the performance of a single-user SR-LDPC code.
Therein, we consider an $(7350, 5888)$ SR-LDPC code of rate $R_{\mathrm{SRLDPC}} \approx 0.80$ that utilizes a $(766, 736)$ non-binary LDPC code over $\mathbb{F}_{256}$.
These parameters were originally selected to facilitate comparisons with similar codes in the literature. 
The performance of this SR-LDPC code is compared to a highly-optimized SPARC/LDPC construction from \cite{greig2017techniques}, a rate $R = 0.8$ NR binary LDPC code with BPSK signalling, and a rate $R = 0.4$ NR LDPC code with BICM using the $4$-PAM constellation.
At $E_b/N_0 = 2.25$~dB, the single-user SR-LDPC code begins to significantly outperform alternate schemes in terms of BER, as seen in the table below.
\begin{table}[h]
\centering
\begin{tabular}{|l|l|l|l|l|}
\hline
$E_b/N_0$ & \multicolumn{1}{|p{12mm}|}{\raggedright SR-LDPC} & \multicolumn{1}{|p{18mm}|}{\raggedright 4-PAM + LDPC BICM}
& \multicolumn{1}{|p{12mm}|}{\raggedright BPSK + LDPC} & \multicolumn{1}{|p{12mm}|}{\raggedright SPARC + LDPC} \\
\hline
1.5 & 0.16397 & 0.18443 & \textbf{0.080139} & -- \tabularnewline
1.75 & 0.12211 & 0.15474 & \textbf{0.069076} & -- \tabularnewline
2.0 & \textbf{0.052382} & 0.091709 & 0.055831 & 0.225 \tabularnewline
2.25 & \textbf{0.004036} & 0.012070 & 0.037056 & -- \tabularnewline
2.5 & \textbf{0.0000287} & 0.000120 & 0.012483 & 0.2 \tabularnewline
\hline
\end{tabular}
\end{table}

Moving forward, we fix the SNR to be $E_b/N_0 = 2.25$~dB and study the performance of the MU-SR-LDPC code obtained when all users employ the aforementioned $(766, 736)~\mathbb{F}_{256}$ LDPC code. 
A crucial parameter for any MU-SR-LDPC scheme is the number of channel uses employed; pragmatically, this parameter is determined by the height of matrix $\Phim$. 
Having implicitly assumed that all users employ the same rate, we define the sum rate to be $R_{\mathrm{sum}} = \frac{K\times 5888}{n_K}$, where $n_K$ is the number of channel uses employed when there are $K$ active users.
We are interested in characterizing the BER performance of this MU-SR-LDPC scheme as a function of $R_{\mathrm{sum}}$.
As a benchmark, we consider a generic orthogonal multiple access (OMA) scheme such as TDMA or FDMA.
In this orthogonal scheme, $n_K = nK$, and each user experiences the exact same performance as reported in~\cite{ebert2023srldpc}, regardless of $K$. 
However, MU-SR-LDPC coding is a \textit{non-orthogonal} multiple access (NOMA) scheme; thus, it is not necessary that $n_K \geq nK$. 

\begin{figure}[tb!]
    \centering
    \begin{tikzpicture}

\definecolor{customred}{rgb}{0.63529,0.07843,0.18431} % red
\definecolor{customblue}{rgb}{0.00000,0.44706,0.74118} % blue
\definecolor{customgreen}{rgb}{0.00000,0.49804,0.00000} % dark green

\begin{semilogyaxis}[
    font=\small,
    width=7cm,
    height=5cm,
    scale only axis,
    every outer x axis line/.append style={white!15!black},
    every x tick label/.append style={font=\color{white!15!black}},
    xmin=0.76,
    xmax=0.96,
    xtick = {0.6, 0.64, ..., 0.96},
    xlabel={$R_{\mathrm{sum}}$},
    xmajorgrids,
    every outer y axis line/.append style={white!15!black},
    every y tick label/.append style={font=\color{white!15!black}},
    ymin=0.001,
    ymax=0.3,
    ytick = {0.001, 0.01, 0.1, 1.0},
    ylabel={BER},
    ymajorgrids,
    yminorgrids,
    legend style={at={(0,1)},anchor=north west, draw=black,fill=white,legend cell align=left}
]

\addplot [
    color=orange,
    solid,
    line width=2.0pt,
    mark size=2.5pt,
    mark=diamond,
    mark options={solid}
]
table[row sep=crcr]{
    0.9601 0.202517 \\
    0.9401 0.178848 \\
    0.9201 0.157050 \\
    0.9000 0.119094 \\
    0.8801 0.084191 \\
    0.8601 0.051808 \\
    0.8401 0.024255 \\
    0.8201 0.009016 \\
    0.8000 0.003524 \\
    0.7801 0.001599 \\
    0.7600 0.000638 \\
};
\addlegendentry{OMA};

\addplot [
    color=customgreen,
    solid,
    line width=2.0pt,
    mark size=2.5pt,
    mark=square,
    mark options={solid}
]
table[row sep=crcr]{
    0.9601 0.200603 \\
    0.9400 0.185457 \\
    0.9200 0.153310 \\
    0.9000 0.113672 \\
    0.8801 0.070489 \\
    0.8600 0.028430 \\
    0.8400 0.011506 \\
    0.8201 0.002444 \\
    0.8000 0.000483 \\
    0.7800 0.000125 \\
};
\addlegendentry{$2$~Users};

\addplot [
    color=customblue,
    solid,
    line width=2.0pt,
    mark size=3.5pt,
    mark=asterisk,
    mark options={solid}
]
table[row sep=crcr]{
    0.9600 0.201184 \\
    0.9400 0.181955 \\
    0.9200 0.158139 \\
    0.9000 0.123645 \\
    0.8800 0.069557 \\
    0.8600 0.021871 \\
    0.8400 0.004940 \\
    0.8200 0.000408 \\
    0.8000 0.000045 \\
};
\addlegendentry{$4$~Users};

\addplot [
    color=customred,
    solid,
    line width=2.0pt,
    mark size=2.5pt,
    mark=triangle,
    mark options={solid}
]
table[row sep=crcr]{
    0.9600 0.203967 \\
    0.9400 0.186160 \\
    0.9200 0.159845 \\
    0.9000 0.125161 \\
    0.8800 0.073134 \\
    0.8600 0.018304 \\
    0.8400 0.001218 \\
    0.8200 0.000004 \\
};
\addlegendentry{$8$~Users};

\addplot [
    color=black,
    dashed,
    line width=2.0pt
]
table[row sep=crcr]{
    0.9417597690118643 0.000001 \\
    0.9417597690118643 1.0 \\
};
\addlegendentry{$C_{\mathrm{sum}}$};

\end{semilogyaxis}
\end{tikzpicture}
    \caption{This plot shows the performance of the MU-SR-LDPC scheme as a function of the sum rate for a varying number of users at $E_b/N_0 = 2.25$~dB.}
    \label{fig:MU-SR-LDPC}
\end{figure}
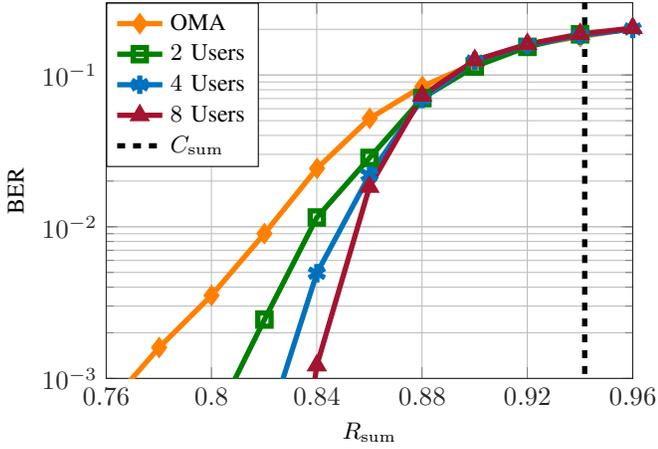

In Fig.~\ref{fig:MU-SR-LDPC}, we plot the BER as a function of $R_{\mathrm{sum}}$ for various values of $K$. 
For convenience, we have also plotted an upper bound on the sum capacity, which is provided below
\begin{equation}
    \begin{split}
        C_{\mathrm{sum}} &< \frac{1}{2}\log\left(1 + K\frac{L}{nK\sigma^2}\right) = \frac{1}{2}\log\left(1 + \frac{L}{n\sigma^2}\right).
    \end{split}
\end{equation} 

From Fig.~\ref{fig:MU-SR-LDPC}, it is clear that, when $E_b/N_0 = 2.25$~dB, the maximum $R_{\mathrm{sum}}$ at which one may achieve a BER of $10^{-3}$ grows with $K$ and is higher than when an orthogonal scheme is utilized.
% This is slightly surprising.
It appears that, with a same set of devices, the system performs better when the users share the entire space compared to scenarios where they would be separated (OMA).
This can be explained, partly, by the fact that orthogonal schemes are not optimal for finite block-lengths.
% Of course, the gains are limited; operations of the MAC are governed by information theoretic limits.
Nevertheless, the fact that performance improves slightly is a strong point for a pragmatic scheme where complexity scales as $\Ocal\left(K\right)$ (per AMP iteration).
The gains can be explained by the fact that, as the number of users increase, the system is shifting away from the effects of finite-block length.
Further, with growth, the actual operation of the system may be getting closer to its asymptotic behavior.
A more precise explanation for this phenomenon remains unclear.
The implications of these empirical observations, on the other hand, are very promising for MU-SR-LDPC codes.
They appear to offer a credible pathway for the joint decoding of MAC users who are transmitting coherently, with reasonable complexity and excellent performance.

\section{Cell-Free MU-SR-LDPC}
\label{section:cell_free_system_model}

\begin{figure}
    \centering
    \begin{tikzpicture}

\definecolor{customred}{rgb}{0.63529,0.07843,0.18431} % red
\definecolor{customblue}{rgb}{0.00000,0.44706,0.74118} % blue
\definecolor{customgreen}{rgb}{0.00000,0.49804,0.00000} % dark green

\begin{semilogyaxis}[
    font=\small,
    width=7cm,
    height=5cm,
    scale only axis,
    every outer x axis line/.append style={white!15!black},
    every x tick label/.append style={font=\color{white!15!black}},
    xmin=-1,
    xmax=2.0,
    xtick = {-1, -0.5, ..., 2.0},
    xlabel={$E_b/N_0$ (dB)},
    xmajorgrids,
    every outer y axis line/.append style={white!15!black},
    every y tick label/.append style={font=\color{white!15!black}},
    ymin=0.001,
    ymax=0.5,
    ytick = {0.00001, 0.0001, 0.001, 0.01, 0.1, 1.0},
    ylabel={BER},
    ymajorgrids,
    yminorgrids,
    legend style={at={(0,0)},anchor=south west, draw=black,fill=white,legend cell align=left}
]

\addplot [
    color=customred,
    solid,
    line width=2.0pt,
    mark size=2.0pt,
    mark=triangle,
    mark options={solid}
]
table[row sep=crcr]{
    -1.0 0.21848505 \\
    -0.75 0.19877378 \\
    -0.5 0.18118207 \\
    -0.25 0.16334985 \\
    0.0 0.14311861 \\
    0.25 0.124262341 \\
    0.5 0.10372028 \\
    0.75 0.08353516 \\
    1.0 0.06465212 \\
    1.25 0.04693699 \\
    1.5 0.02364515 \\
    1.75 0.00290613 \\
    2.0 0.0000373093443 \\
};
\addlegendentry{SU};

\addplot [
    color=customgreen,
    solid,
    line width=2.0pt,
    mark size=2.0pt,
    mark=square,
    mark options={solid}
]
table[row sep=crcr]{
    -1.0 0.26506793 \\
    -0.75 0.24954484 \\
    -0.5 0.22666101 \\
    -0.25 0.20066403 \\
    0.0 0.15367547 \\
    0.25 0.124704201 \\
    0.5 0.10374151 \\
    0.75 0.08342589 \\
    1.0 0.06479535 \\
    1.25 0.04658854 \\
    1.5 0.02381997 \\
    1.75 0.00314599 \\
    2.0 0.0000298584327 \\
};
\addlegendentry{UE $0$};

\addplot [
    color=customblue,
    solid,
    line width=2.0pt,
    mark size=2.0pt,
    mark=asterisk,
    mark options={solid}
]
table[row sep=crcr]{
    -1.0 0.1161447 \\
    -0.75 0.09615489 \\
    -0.5 0.06988111 \\
    -0.25 0.04862665 \\
    0.0 0.01072071 \\
    0.25 0.0000560461957 \\
    0.5 0.0 \\
    0.75 0.0 \\
    1.0 0.0 \\
    1.25 0.0 \\
    1.5 0.0 \\
    1.75 0.0 \\
    2.0 0.0 \\
};
\addlegendentry{UE $1$};

\addplot [
    color=orange,
    solid,
    line width=2.0pt,
    mark size=2.0pt,
    mark=asterisk,
    mark options={solid}
]
table[row sep=crcr]{
    -1.0 0.26608696 \\
    -0.75 0.25005774 \\
    -0.5 0.22434783 \\
    -0.25 0.19829297 \\
    0.0 0.15402314 \\
    0.25 0.124313293 \\
    0.5 0.103262 \\
    0.75 0.08341287 \\
    1.0 0.06517691 \\
    1.25 0.04711702 \\
    1.5 0.02293954 \\
    1.75 0.00277853 \\
    2.0 0.0000307350105 \\
};
\addlegendentry{UE $2$};

\end{semilogyaxis}
\end{tikzpicture}
    \caption{This shows the performance of the MU-SR-LDPC scheme applied to a cell-free setting.
    There are two base stations and three users; the connectivity graph is $\mathcal{K}_1 = \{0, 1\}$ and $\mathcal{K}_2 = \{1,2\}$.
    The performance of a single user (SU) in a single cell network with identical parameters is provided as a benchmark.}
    \label{fig:Cell-Free}
\end{figure}
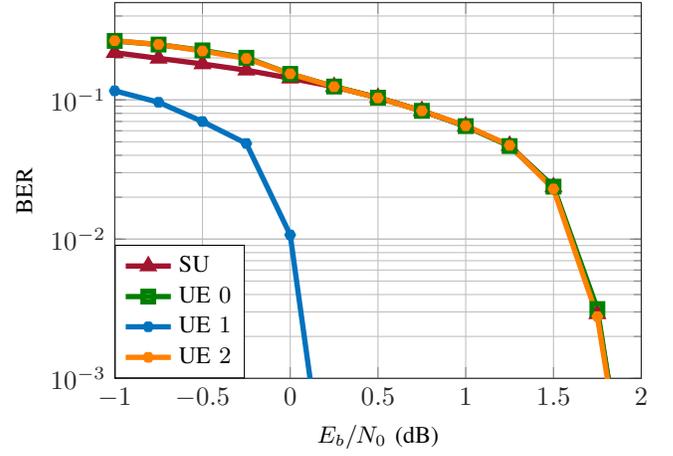

The extension of the MU-SR-LDPC scheme to the cell-free setting can be explained in a straightforward manner.
We consider a simple model where each base station receives signals from a subset of the user population in the form of an observation vector.
This can be called the bipartite model because every base station is connected to a subset of users, and every user is linked to a subset of the base stations.
The signal received at access point~$b$ is given by
\begin{equation}
\textstyle  \yv_b = \sum_{k \in \mathcal{K}_b} \Am_k \sv_k + \zv_b,
\end{equation}
where $\mathcal{K}_b$ is the set of graph neighbors of base station~$b$.
The AMP-BP algorithm progresses in a manner akin to the single-cell case, except for two critical modifications.
First, every base station must compute its own residual vector
\begin{equation}
\textstyle  \zv_b^{(t)} = \yv_b - \sum_{k \in \mathcal{K}_b} \hat{\yv}_k^{(t)} .
\end{equation}
Second, a user may appear in many residuals, $\big\{ \zv_b^{(t)} : b \in \mathcal{B}_k \big\}$, where $\mathcal{B}_k$ denotes the graph neighborhood of user~$k$.
When this is the case, the cell-free back-end must compute and combine several effective observations
\begin{equation}
\rv_{b,k}^{(t)} = \Am_k^{\mathrm{T}} \zv_b^{(t)} + \sv_k^{(t)} \quad b \in \mathcal{B}_k .
\end{equation}
Since every effective observation is approximately equal to true signal $\sv_k$ embedded in i.i.d.\ Gaussian noise of known variance, one can form a master effective observation through optimal combining 
(see proof in Appendix~\ref{appendix:OptimalCombining})
\begin{equation*}
\rv_{k}^{(t)} = \sum_{b \in \mathcal{B}_k} \frac{\rv_{b,k}^{(t)}}{\sum_{b' \in \mathcal{B}_k}\left(\frac{\tau_b^2}{\tau_{b'}^2}\right)} .
\end{equation*}
Once the master effective observation is obtained in lieu of \eqref{eq:EffectiveObservation-Plus}, the state estimate $\sv_{k}^{(t+1)}$ and the estimated contributions $\hat{\yv}_k^{(t)}$ are computed as before, see \eqref{eq:EstimatedContributions} and \eqref{eq:AMP-Denoising-Plus}.
The AMP-BP algorithm then proceeds as originally described.
For a rudimentary cell-free scenario with three users and two base stations, the proposed approach performs remarkably well, as illustrated in Fig.~\ref{fig:Cell-Free}.
It is pertinent to note that the joint decoding process for a cell-free scenario is intrinsically local.
During every round, access point~$b$ only needs information from connected users in $\mathcal{K}_b$.
Likewise, the computation for user~$k$ only requires residuals from neighboring access points in $\mathcal{B}_k$.
The computations then take the familiar form of message passing over a sparse bipartite graph~\cite{kschischang2001factorgraph,loeliger2004introduction}.

\section{Conclusion}

This article proposes a novel coding scheme named MU-SR-LDPC coding for non-orthogonal multiple access (NOMA) and cell-free systems.
This scheme is suitable for the application of the AMP-BP algorithm, with appropriate modifications.
This yields a low-complexity solution whose performance in both settings is excellent.
The structure of the coding scheme and its processing are closely connected to several recent contributions in information theory on AMP applied to wireless systems.
Nevertheless, the proposed concatenated structure that pairs an LDPC outer code to a sparse regression inner code is new for coordinated multi-user communication channels.
Future directions for this work include theoretical pursuits related to capacity and state evolution.
Pragmatically, phase uncertainty, timing offset, complex signaling, and MIMO settings are interesting future considerations.

\newpage

\bibliographystyle{IEEEbib}
\bibliography{IEEEabrv, isit2024}

\newpage
\appendices
\section{Proof of Optimal Combining}
\label{appendix:OptimalCombining}

\begin{proposition}
Let $\cv \in \mathbb{R}^{B}$ denote the optimal weights to be used in combining $B$ effective observations. 
Then, each entry $c_j \in \cv$ is given by
\begin{equation}
c_j = \frac{1}{\sum_{i \in[B]} \left( \frac{\tau_j^2}{\tau_i^2} \right)} .
\end{equation}
\end{proposition}
\begin{IEEEproof}
We begin by defining what constitutes an optimal combining vector. 
First, the vector must sum to $1$, so that properties of the underlying signal are unaffected by the combining operation. 
Second, we would like the effective noise variance after combining to be as small as possible. 
With these objectives in mind, we formulate an optimization problem in standard form: 
\begin{alignat}{2}
\min_{\cv} & & \quad &\sum_{i \in [B]} c_i^2\tau_i^2 \\
\operatorname{s.t.} & & \quad & \|\cv\|_1 = 1 .
\end{alignat}
Then, we define the Lagrangian function as
\begin{equation}
L(\cv, \nu) = \sum_{i \in [B]} c_i^2\tau_i^2 + \nu\left(\|\cv\|_1 - 1\right) .
\end{equation}
Since the problem is convex, the Karush–Kuhn–Tucker (KKT) conditions are sufficient for optimality. 
Thus, it suffices to find a $\cv$ and $\nu$ such that $\|\cv\|_1 = 1$ and
\begin{equation}
\nabla_{\cv} \left[ \sum_{i \in [B]} c_i^2 \tau_i^2 + \nu \left(\|\cv\|_1 - 1\right) \right] = 0.
\end{equation}
In order for the second condition to hold, it must be that
\begin{equation} \label{eq:optimal_mrc_weights_wj}
c_j = \frac{-\nu}{2\tau_j^2}.
\end{equation}
For the first condition to hold, $\|\cv\|_1 = 1$; thus, from this constraint, we can compute $\nu$ as
\begin{equation} \label{eq:optimal_mrc_weights_nu}
\nu = \frac{-2}{\sum_{i \in [B]} \frac{1}{\tau_i^2}}.
\end{equation}
The result follows by substituting \eqref{eq:optimal_mrc_weights_nu} into \eqref{eq:optimal_mrc_weights_wj}.
\end{IEEEproof}

\end{document}